\documentclass[fleqn,twoside]{article}
\usepackage{espcrc2}

\usepackage{graphicx}
\usepackage[figuresright]{rotating}

\newcommand{\AmS}{{\protect\the\textfont2
  A\kern-.1667em\lower.5ex\hbox{M}\kern-.125emS}}

\title{Cost of dynamical quark simulations with improved staggered quarks}

\author{Steven Gottlieb\address{Department of Physics---SW117; Indiana
	University; Bloomington, IN 47405; USA}
	\address{Theory Group MS106; Fermilab; P.O. Box 500;
	Batavia, IL 60510--0500; USA}
        \thanks{At Fermilab until June 15, 2002.}
        }
       
\begin{document}

\begin{abstract}
The cost of dynamical quark simulations with improved staggered quarks
is estimated based on current and planned running by the MILC collaboration.
I find that a few 10s of Teraflop years should be sufficient to calculate
down to a lattice spacing of 0.045 fm.
\vspace{1pc}
\end{abstract}

\maketitle

\section{INTRODUCTION}
Several {\it caveats\/} should be pointed out to the reader.
First, I think that the past 20 years of my life are evidence
of my inability to estimate the time and effort
necessary to calculate the spectrum of QCD.  Used to doing analytic
calculations that took a few months at most, I did not imagine as I started
my first Monte Carlo calculation 
that 20 years later I would still be doing similar calculations.  I was a
postdoc at Fermilab then, and as I write this I am back at Fermilab on 
sabbatical.  With great pleasure I see how far
we have come, and look forward to an exciting future of improved calculations.

Caveat two is that this write-up is not a transcript
of what I said in Berlin.  Caveat three is that
this does not represent a MILC consensus statement.
I did my best to extract from past experience 
what is required for future calculations, but the whole collaboration
has not had an opportunity to check or react.


\section{TIME ESTIMATE}
Since the CG routine is no longer so dominant, the formula for
counting operations is not quite so simple.  For a 2+1 flavor run,
let: 
$\tau =$ \# of time units per independent configuration;
$O_F=$ \# of operations for fermion force per site;
$O_{FL}=$ \# of operations for fat link calculation per site;
$O_{CG}=$ \# of operations per CG iteration per site;
$N_{CG}^s=$ \# of iteration to solve CG for $s$-quark.
Denoting the quark masses $m_s$ and $m_l$, we use a time step $\Delta t =
2/3 m_l$, and we expect $N_{CG}^l = N_{CG}^s \cdot m_s/m_l$.  The operation
count is
\begin{eqnarray}
&={3\over2}{L\over a}^4 \tau {1\over m_l}[ (2 O_F + O_{FL} + O_{CG} N_{CG}^s)
  \nonumber\\
&+ O_{CG} N_{CG}^s {m_s\over m_l}]
\end{eqnarray}

\subsection{Production running}
MILC, already in production with three dynamical quarks,
has completed
runs with lattice spacing 0.2 and 0.13 fm \cite{spectrum}.  
The Asqtad action has leading errors of order $a^2 g^2$ 
and uses tadpole improved coefficients in the action \cite{asqtad}.
A series of runs was done to allow a smooth interpolation
between the quenched approximation and the 2+1 flavor world.  
The coupling is tuned as the quark masses are reduced to fix a length scale
determined from the heavy quark potential \cite{potential}.
We have eight dynamical runs.  In four, there are three degenerate 
quarks with mass 
$8 m_s$, $4 m_s$, $2 m_s$ or $m_s$.  In four runs, the
dynamical strange quark is fixed at $m_s$ and the light
quark mass is 0.8, 0.6, 0.4 or 0.2 times $m_s$.

Runs are $\approx$ 2000--3000 molecular dynamical time units.
One significant issue is how the autocorrelation time scales as
the lattice spacing is reduced.  We cannot yet provide
numerical evidence for this scaling law.  Currently, we are saving every
sixth trajectory.  We see some autocorrelation between successive
lattices.  For analysis, we bin in groups of four, {\it i.e.},
24 time units.

Also important is how many independent configurations are needed to achieve
the desired accuracy.
This number clearly depends on the quark masses and the quantity under
study, but we (the lattice community) may not be aware of
its dependence on such quantities as the action and volume.  As an example
of the latter, MILC has done some extensive tests of finite size effects
in the past.  It is much easier to get accurate masses on large
volumes than on the smaller ones.  Thus, we need more configurations
for the smaller volumes.  In our projects, we often use
big volumes compared to some groups using Wilson or Clover quarks.
We may need fewer independent configurations to achieve the same accuracy.
(Not to mention no finite-size effects.)

Table \ref{tab:requiredflops} has timing estimates for some current
runs at 0.09 fm.
We print timing for the conjugate gradient routine, but
not for other parts of the code.  We run on several different machines,
but this estimate is based on the assumption of a speed of 200 MF/CPU.  
To get the operation count we will assume the entire code is running at
200 MF, not just the CG (the only part regularly timed).
The table contains the number of
node hours required to create a configuration we will store given
the parameters we use for time step and residual.  The lattice volume
is $28^3 \times 96$.  In the first line, the quarks are degenerate, so
we need only one quark field and $N_f=3$.  (For Asqtad, $O_F\approx$ 420K,
$O_{FL}\approx$ 51K, $N_{CG}=1187$ and $N_{CG}^s=236$, giving an opcount
within 30\% of the table value.)

For the other two runs,
we need two quark fields, with $N_f=2$ and 1.  In the last run, we 
reduce the light quark mass by a factor of two compared to the one above it.
Traditional scaling laws would predict a factor of 4 increase in
computation from halving the time step and doubling the number of CG
iterations; however, as the code is no longer dominated by the CG
routine, the time only increases by 2.4.  For the lightest mass run,
we plan to store 400 configurations.  This amounts to 0.145 TF-years. 

Now we attempt to address the issue of what it would take to do a
calculation of the quality of the CPPACS quenched Wilson quark calculation.
CPPACS states that their smallest lattice spacing is 0.05 fm.  However,
they only have 150 configurations there, and their error bars are significantly
larger than at the stronger couplings.  I have always wondered how their
continuum extrapolation would change eliminating either the smallest or
largest $a$.  Is the hardest part of the calculation important in reducing
the final error?
If we halve our current
lattice spacing to get to 0.045 fm, we will be somewhat closer to the
continuum limit than they were.  
Assuming we generate 400 configurations, we can roughly
estimate the time required by multiplying our current light mass run
by a factor of $2^6$ to $2^8$.  The smaller factor assumes four powers from
the volume, one from the time step and another from the CG iterations (which
may be an overestimate as the CG no longer dominates).  The larger factor
allows for a doubling of the autocorrelation time, and doubles the time for
additional runs at heavier masses.  This yields an estimate of 10--40
TF-yr of running.  
CPPACS's next to smallest $a$ was 0.064 fm.  
If we were to go to a lattice spacing of 0.06 fm, the
increase in time from our present run would be from 11 to 26 depending on
how things scale.  This would only require about 2--4 TF-yr.

Despite my first caveat and the surprise of many in the audience, I believe
these are reasonable estimates for the runs outlined.
MILC has not had a dedicated computer, but we have
been able to do significant calculations partly because staggered quarks do
not require as much computation as Wilson/clover.

\begin{table}[t]
\caption{Computational requirement of runs}
\label{tab:requiredflops}
\begin{tabular}{ccc}
\hline 
$m_{u,d}$ & node-hr/conf & flops ($10^{14}$/conf) \\
\hline 
$m_s$ & 900 & 6.5 \\
$0.4 m_s$ & 4096 & 29.5 \\
$0.2 m_s$ & 9780 & 70.4 \\
\end{tabular}
\end{table}

\end{document}